\documentclass[twocolumn]{aastex62}

\usepackage{amsmath}

\received{-}
\revised{-}
\accepted{-}
\submitjournal{ApJ}

\shorttitle{}
\shortauthors{Dr\c{a}\.{z}kowska \& Szul\'agyi}
\begin{document}

\title{DUST EVOLUTION AND SATELLITESIMAL FORMATION IN CIRCUMPLANETARY DISKS}

\correspondingauthor{Joanna Dr\c{a}\.{z}kowska}
\email{joanna.drazkowska@lmu.de}

\author[0000-0002-9128-0305]{Joanna Dr\c{a}\.{z}kowska}
\affil{Institute for Computational Science, University of Zurich, Winterthurerstrasse 190, 8057 Zurich, Switzerland}
\affil{University Observatory, Faculty of Physics, Ludwig-Maximilians-Universit\"{a}t M\"{u}nchen, Scheinerstr. 1, 81679 Munich, Germany}

\author[0000-0001-8442-4043]{Judit Szul\'{a}gyi}
\affil{Institute for Computational Science, University of Zurich, Winterthurerstrasse 190, 8057 Zurich, Switzerland}
%\affil{Institute for Particle Physics and Astrophysics, ETH Zurich, Wolfgang-Pauli-Strasse 27, 8093, Zurich, Switzerland}

%% Note that the \and command from previous versions of AASTeX is now
%% depreciated in this version as it is no longer necessary. AASTeX 
%% automatically takes care of all commas and "and"s between authors names.

%% AASTeX 6.2 has the new \collaboration and \nocollaboration commands to
%% provide the collaboration status of a group of authors. These commands 
%% can be used either before or after the list of corresponding authors. The
%% argument for \collaboration is the collaboration identifier. Authors are
%% encouraged to surround collaboration identifiers with ()s. The 
%% \nocollaboration command takes no argument and exists to indicate that
%% the nearby authors are not part of surrounding collaborations.

%% Mark off the abstract in the ``abstract'' environment. 

\begin{abstract}

It is believed that satellites of giant planets form in circumplanetary disks. Many of the previous contributions assumed that their formation process proceeds similarly to rocky planet formation, via accretion of the satellite seeds, called satellitesimals. However, the satellitesimal formation itself poses a nontrivial problem as the dust evolution in circumplanetary disk is heavily impacted by fast radial drift and thus dust growth to satellitesimals is hindered. To address this problem, we connected state-of-the art hydrodynamical simulations of a circumplanetary disk around a Jupiter-mass planet with dust growth and drift model in a post-processing step. We found that there is an efficient pathway to satellitesimal formation if there is a dust trap forming within the disk. Thanks to natural existence of an outward gas flow region in the hydrodynamical simulation, a significant dust trap arises at the radial distance of 85~R$_{\rm J}$ from the planet, where the dust-to-gas ratio becomes high enough to trigger streaming instability. The streaming instability leads to efficient formation of the satellite seeds. Because of the constant infall of material from the circumstellar disk and the very short timescale of dust evolution, the circumplanetary disk acts as a satellitesimal factory, constantly processing the infalling dust to pebbles that gather in the dust trap and undergo the streaming instability.

\end{abstract}

%% Keywords should appear after the \end{abstract} command. 
%% See the online documentation for the full list of available subject
%% keywords and the rules for their use.

\section{Introduction}

The Galilean moons of Jupiter were the first bodies found to orbit a planet other than Earth. They are among the largest moons in the Solar System. With their almost circular and aligned orbits, they are believed to have formed in a disk surrounding the young Jupiter \citep[see, e.g.,][]{1982Icar...52...14L}. Although many models of such disk were built in the past \citep[see, e.g.,][]{1999ApJ...526.1001L, 2002AJ....124.3404C, 2003Icar..163..198M, 2005A&A...439.1205A, 2009MNRAS.397..657A, 2013ApJ...767...63S, 2014ApJ...785..101F, 2017AJ....153..194F, 2016MNRAS.460.2853S, 2017MNRAS.464.3158S}, formation of the Galilean satellites is still a subject of intense research.

Majority of the previous works focused on the Galilean moon formation with their gravitationally bound precursors as a starting point, called satellitesimals \citep[see, e.g.][]{2010ApJ...714.1052S, 2012ApJ...753...60O, 2016Icar..266....1M, 2018MNRAS.475.1347M}. However, the formation of these satellitesimals was not explained in a convincing way. Recently, \citet{2017ApJ...846...81S} showed that this is in fact a tough problem, as the conditions in the circumplanetary disk lead to a very fast radial drift and particle sticking to satellitesimals sizes is possible only in rare setups. 

The problem of satellitesimal formation in a circumplanetary disk is to a large degree analogical to the problem of planetesimal formation in a circumstellar (also called protoplanetary) disk. Particle-gas interactions determine the redistribution of solids at essentially all particle sizes \citep{2014prpl.conf..411T}. These interactions do also determine random velocities between particles of different sizes, which drive their collisions \citep{2016SSRv..205...41B}. The outcomes of dust-aggregate collisions are well studied in laboratory experiments. While low-velocity collisions result in sticking of small particles, the impact speeds increase with aggregate size, such that further collisions result in bouncing, erosion or fragmentation of millimeter to centimeter-sized aggregates \citep{2010A&A...513A..56G, 2016ApJ...827..110K, 2017ApJ...834..145B}. Possibilities of direct growth to kilometer-sized objects are further reduced by the short timescale of radial drift, which is caused by the loss of angular momentum due to particles interactions with the sub-Keplerian gas disk. The radial drift timescale is expected to be even shorter in disks around planets and low mass stars than in disks around solar mass stars \citep{2013A&A...554A..95P, 2018MNRAS.479.1850Z}, due to their smaller radial extent and higher divergence from the Keplerian rotation.

A generally accepted solution to the fragmentation and drift barriers is the streaming instability \citep{2007Natur.448.1022J}. This process leads to a spontaneous clumping of dust into filaments. Some of these filaments become massive enough to undergo collapse and form gravitationally bound bodies. However, for the streaming instability to work, dust grains must be large enough and the local solids-to-gas ratio must be sufficiently enhanced \citep{2010ApJ...722.1437B, 2014A&A...572A..78D, 2015A&A...579A..43C}. In this paper, we show that the streaming instability is also able to form satellitesimals in a disk around proto-Jupiter. The radial drift is stopped thanks to an outward gas flow seen in viscosity included hydrodynamical models \citep{2010MNRAS.405.1227M,2012ApJ...747...47T,2014ApJ...782...65S, 2016ApJ...832..105F,2017ApJ...842..103S,2017MNRAS.464.3158S} and therefore a dust trap region is easily created where particles can grow to pebbles, which can then undergo the streaming instability.

In a corresponding paper, \citet{2018MNRAS.480.4355C} used the results of our work to perform population synthesis models on the satellite formation around Jupiter. They found that it is indeed possible to reproduce the Galilean satellite-configuration, preferably given a long-lived circumplanetary disk with a high solids-to-gas ratio. The satellite generations form in a sequence after each other and most of them are lost into the planet due to fast radial migration. The ice-rich satellites that we see today have likely formed very late during the disk evolution, when it was depleted in gas and cold enough to sustain water ice.

This paper is organized as follows. Section~\ref{disk} describes our circumplanetary disk model extracted from the simulations of \citet{2017ApJ...842..103S} and typical velocities and timescales determining dust evolution in this disk. Section~\ref{methods} outlines the numerical methods that we have used to model dust evolution and satellitesimal formation. Section~\ref{results} presents the results of our research, which are then discussed in Section~\ref{discussion} and summarized in Section~\ref{summary}.

\section{The circumplanetary disk}\label{disk}

In this section, we describe the circumplanetary disk (hereafter CPD) model used in this paper as well as discuss its properties that are important for dust evolution. 

\subsection{Hydrodynamical simulations}\label{hydro}

Our CPD model is extracted from one of the numerical simulations presented by \citet{2017ApJ...842..103S}, corresponding a relatively late, evolved state of the circumjovian disk. 
The 3-D hydrodynamic simulation was performed with the JUPITER code \citep{2016MNRAS.460.2853S}, which is a nested mesh Godunov algorithm that solves the basic hydrodynamical equations, including the total energy equation and the flux limited diffusion approximation \citep{1989A&A...208...98K,2011A&A...529A..35C}. This means that the gas temperature is realistically calculated in each cell, under the heating mechanisms (adiabatic compression e.g.~from the accretion process, viscous heating) and the cooling processes (adiabatic expansion, radiative dissipation). The simulation applied a constant kinematic viscosity with a value of $1.02\cdot10^{20}$~cm$^2$~s$^{-1}$. The adiabatic index was set to 1.43, while the mean molecular weight was corresponding to 2.3. The temperature at each cell was calculated with the use of a gas-dust opacity table of \citet{1994ApJ...427..987B} with the assumption of 1\% constant dust-to-gas ratio. Therefore, even though the dust component is not explicitly simulated, its contribution to the temperature is taken into account. The planet was included through a 1 Jupiter mass (M$_{\rm J}$) point mass, with a temperature of 2000~K, which corresponds to the forming Jupiter at approx 1-3~Myrs, depending on evolutionary models \citep{1995ApJ...450..463G,2017A&A...608A..72M}. The simulation contained an 11~M$_{\rm J}$ circumstellar disk ranging between 2.08 AU to 12.40 AU, where the planet (Jupiter) was placed at 5.2 AU. The nested meshes allowed us to have a sub-planetary resolution in the CPD, and our smallest cell-diagonal corresponds to $\sim80$\% of the Jupiter diameter, i.e.~approx. 112,000 km.

\begin{figure}
\plotone{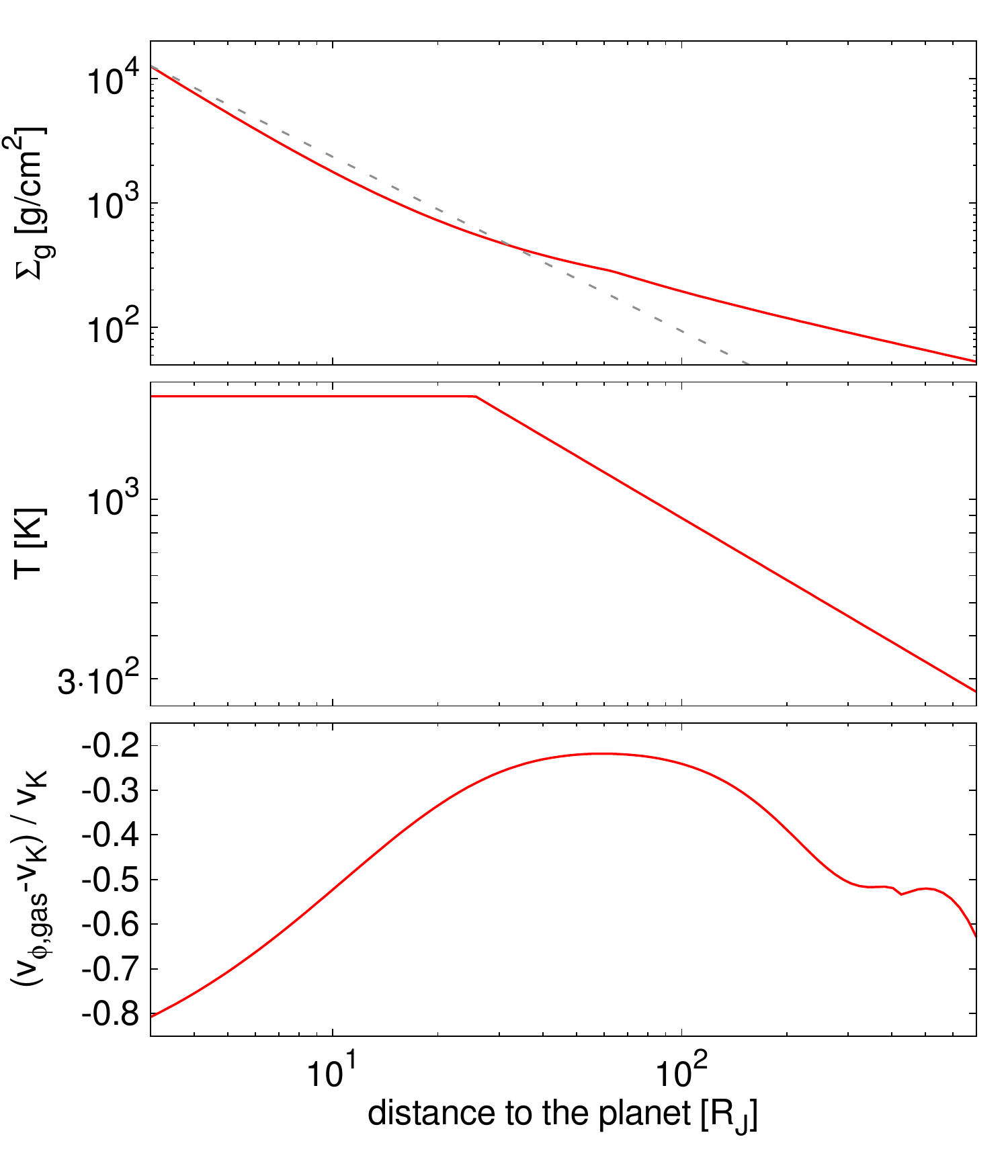}
\caption{The azimuthally-averaged CPD structure derived from the hydrodynamical simulations and used in this paper. {\it Upper panel}: Vertically integrated density of gas. $r^{-1.4}$ profile is displayed with the dashed line for reference. {\it Middle panel:} Midplane temperature of the gas. {\it Bottom panel:} Difference between gas and Keplerian rotation, normalized by the local Keplerian velocity. Values close to zero indicate Keplerian rotation, lower values indicate sub-Keplerian rotation.} \label{fig:disk}
\end{figure}

Figure~\ref{fig:disk} presents the basic characteristics of the CPD model used throughout this paper. The surface density, showed in the upper panel, is the highest towards the inner edge of the disk and decreases outwards. The decrease roughly corresponds to a power law of $\Sigma_{\rm g}\propto r^{-1.4}$ used by \citet{2018MNRAS.480.4355C}. The gas temperature in the midplane, displayed in the middle panel, corresponds to $T\propto^{-0.6}$, except for the inner part of the planet where we limit it to 2000~K corresponding to the late, forming Jupiter's effective temperature. The bottom panel shows the difference between rotation velocity of gas and the Keplerian velocity. The reason why the gas is so remarkably sub-Keplerian is the pressure-support, which is stronger when the disk is hotter.

\begin{figure}
\plotone{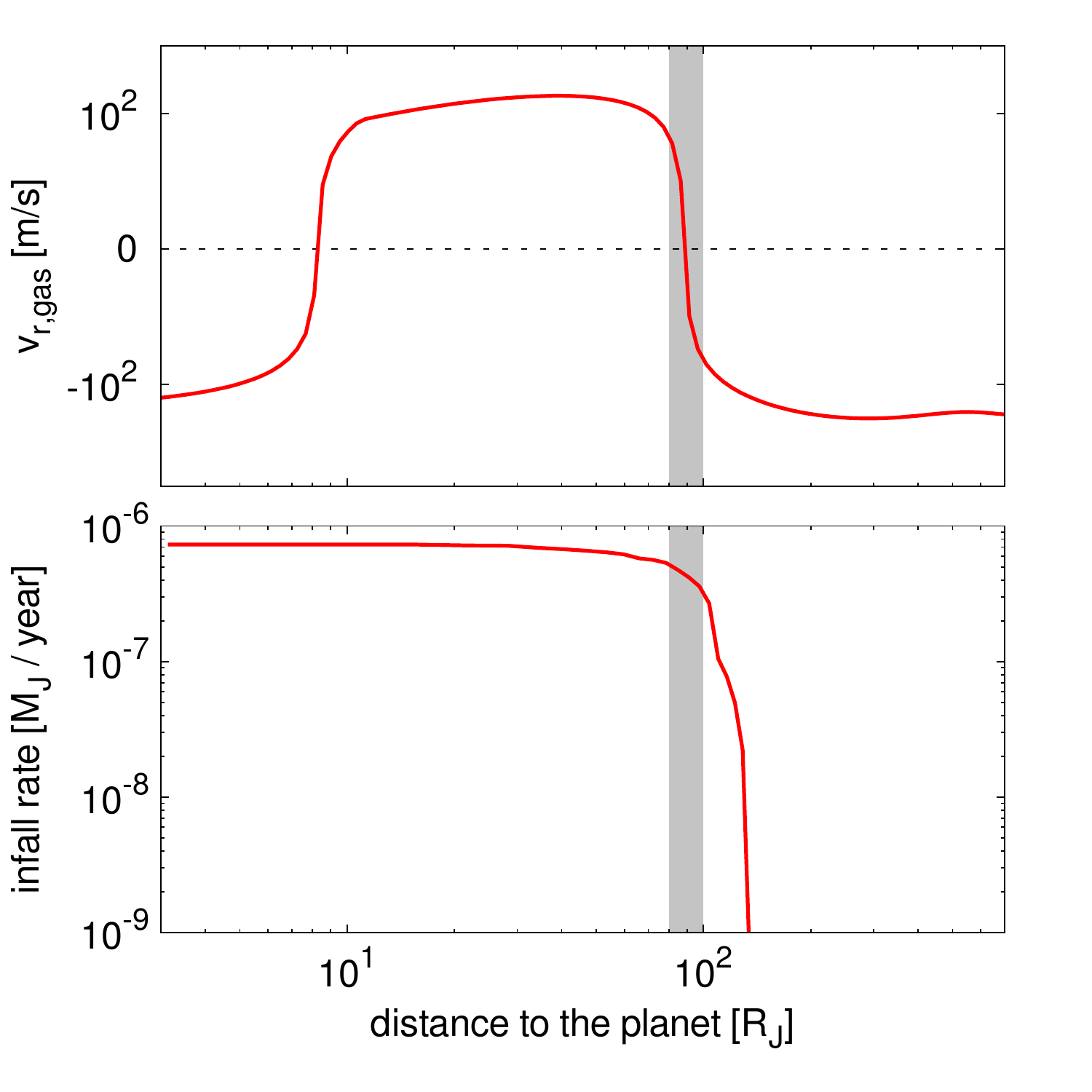}
\caption{{\it Upper panel}: Radial velocity of gas in the midplane of the CPD. The shaded region marks where the gas velocity changes direction from outward ($v_{\rm{gas}}>0$) to inward ($v_{\rm{gas}}<0$). {\it Bottom panel:} Infall onto the CPD extracted from the hydrodynamical simulations and used in some of the models.}\label{fig:infall}
\end{figure}

Although the characteristics displayed in Figure~\ref{fig:disk} may suggest that the CPD is to a large degree analogical to a standard circumstellar disk setup, there are two aspects that make it in fact a very different environment. These are presented in Figure~\ref{fig:infall}. The upper panel of Figure~\ref{fig:infall} shows the midplane radial velocity of gas. The negative values correspond to inward and the positive values to outward flow. A large fraction of the CPD midplane, between approximately 10 and 100 Jupiter radii (R$_{\rm J}$), flows outwards. This is caused by the combination of viscous stress and pressure due to the hot planet, which together enforces the outward flow, in competition with shocks generated by the vertical influx, which make the gas lose its angular momentum and flow inwards. \citet[][their Section 3.3]{2014ApJ...782...65S} showed that even a small change in the viscosity can turn the gas flow from inward to outward (see also their Figures 9 and 10). Subsequently, the gas flows back to the circumstellar disk, and so-called meridional circulation pattern is formed. Similar result was also found by another authors \citep[see, e.g.][]{2012ApJ...747...47T}.

As mentioned before, unlike a circumstellar disk, the CPD is not a closed reservoir of mass, as it is continuously fed by the new gas-dust material from the vertical direction from the circumstellar disk (with a rate of $2\cdot10^{-6}$~M$_{\rm J}$~per year in this simulation), through the planetary gap \citep{2014ApJ...782...65S}. This infall happens mainly to the inner part of the CPD, as plotted in the bottom panel of Figure~\ref{fig:infall}. As we will show later, this properties do completely change the evolution of dust in the CPD with respect to a circumstellar disk, as the outward gas flow is strong enough to save dust aggregates from falling onto the planet.

In our simulations, the dust trap (i.e.~the location where the radial velocity of gas changes sign, see Section~\ref{dust}) is located at about $85$~R$_{\rm J}$ from the planet. However, this exact value is very sensitive to the hydrodynamical simulation setup, namely the planet mass, the gas viscosity, the heating/cooling processes (which determine how pressure-supported is the disk), as well as the opacity and the local optical depth. Therefore, we note that the location of the dust trap can and will vary with different CPD setups, but that would not change the main results presented in this paper, as the importance is to have a dust trap within the CPD in a first place and the significance of its location is secondary.
 
\subsection{Dust evolution pattern}\label{dust}

The dust component was not explicitly included in the original hydrodynamical simulation of \citet{2017ApJ...842..103S}. Therefore, in this subsection, we focus on what the setup obtained in the gas simulation means for dust evolution. 

\begin{figure}
\plotone{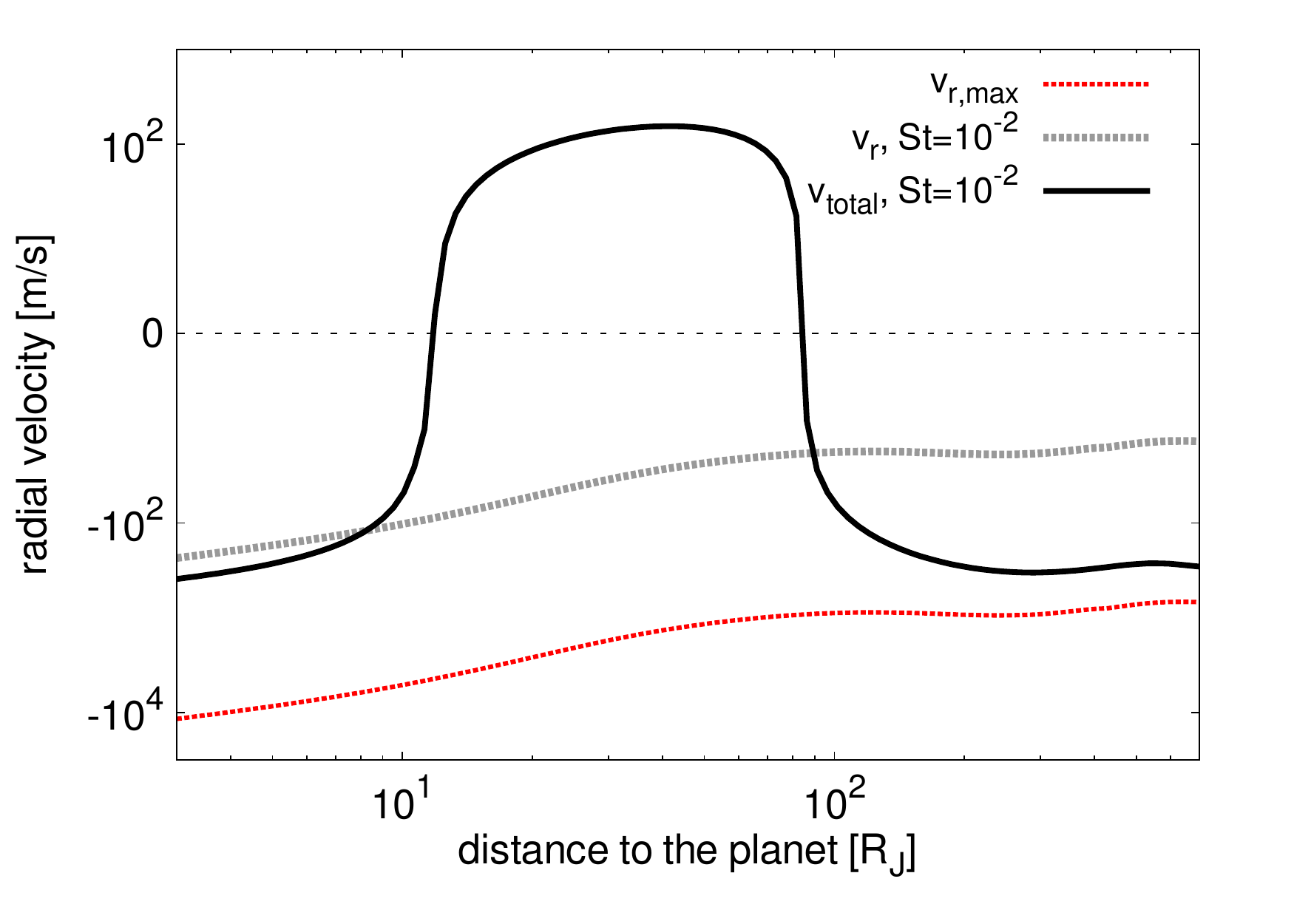}
\caption{Radial velocities in the disk midplane that determine dust redistribution: the nominal maximum radial drift velocity $v_{\rm{r,max}}$ (red dotted line), the radial drift velocity of pebbles with ${\rm{St}}=10^{-2}$ (gray dotted line) and their total advection speed $v_{\rm{total}}$ taking into account the advection by gas (black solid line). The negative radial velocity values indicate inward drift and the positive values indicate outward drift.} \label{fig:vels}
\end{figure}

As in a circumstellar disk, the gas rotation velocity $v_{\phi,\rm{gas}}$ in the CPD is sub-Keplerian. The difference between the Keplerian rotation and gas is typically parametrized by $\eta$:
\begin{equation}\label{eta}
\eta \approx \frac{v_{\phi,\rm{gas}}-v_{\rm K}}{v_{\rm K}},
\end{equation}
where $v_{\rm K}$ is the Keplerian velocity. While in a standard circumstellar disk this difference is less than 1\% \citep{2007astro.ph..1485A}, in our CPD model it is typically 20\%, and can be as high as 80\% (see the bottom panel of Figure~\ref{fig:disk}). This means that the dust grains, which would normally follow the Keplerian rotation, feel a very strong headwind and thus should lose their angular momentum and quickly fall onto the planet with a maximum velocity of $v_{\rm{r,max}}=\eta v_{\rm K}$. We plot the nominal maximum radial drift velocity $v_{\rm{r,max}}$ with red dotted line in Figure~\ref{fig:vels}. It is on the order of tens kilometers per second. This velocity could only be reached by grains with dimensionless stopping time (also called Stokes number, $St$) corresponding to unity, which in the CPD environment translates into the physical size of tens of centimeters. As we will explain later, it is unlikely that such large grains grow in this disk, so it makes sense to also plot the drift velocity for smaller grains, in this case with $St=10^{-2}$, which is two order of magnitude lower (gray dotted line), because the radial drift velocity scales with dimensionless stopping time $St$ as
\begin{equation}\label{vdrift}
v_{\rm r} = \frac{2\ St\ v_{\rm{r,max}}}{1+ St^2}.
\end{equation}

As discussed in Section~\ref{hydro}, \citet{2017ApJ...842..103S} found that the CPD has a wide region where the gas flows outwards with velocities as high as hundreds meters per second (see the upper panel of Figure~\ref{fig:infall}), which magnitude is comparable to the inward drift velocity of pebbles. As the dust grains are coupled to the gas by aerodynamic drag force, this gas flow results in an additional component of dust radial velocity
\begin{equation}\label{vadv}
v_{\rm{r,adv}} = \frac{v_{\rm{r,gas}}}{1+ St^2},
\end{equation}
where $v_{\rm{r,gas}}$ is the radial velocity of gas flow in the midplane.

When we add up the two contributions: the radial drift caused by the loss of angular momentum due to the headwind and the advection caused by the coupling of grains to the flowing gas (black solid line in Figure~\ref{fig:vels}), we find that the pebbles should be saved from the radial drift as their total velocity is outward in a large part of the CPD.
 
\begin{figure}
\plotone{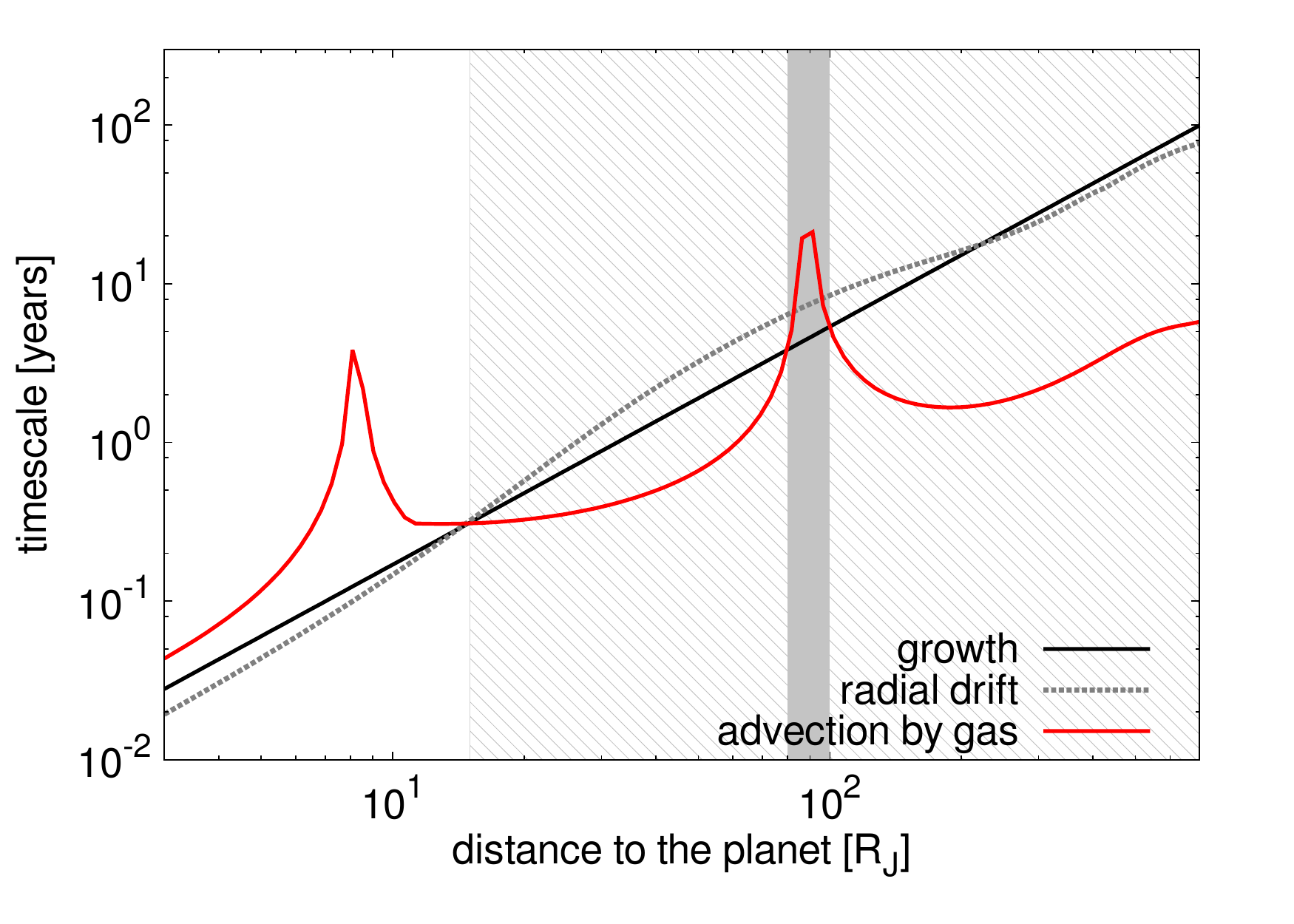}
\caption{Comparison of dust evolution timescales in different parts of the CPD. The inner part of the disk is dominated by the radial drift, while the outer part is dominated by the advection by gas flow (the hatched area). There is a narrow region where the dust growth timescale is the shortest of the three (the shaded region), which corresponds to the location where the gas flow velocity changes direction (see Figure \ref{fig:infall}).} \label{fig:times}
\end{figure}

To put the radial velocities into context, we plot the timescale over which a pebble would be lost into the planet due to the radial drift with gray dotted line in Figure~\ref{fig:times}. This is on the order of days for the inner part of the disk and hundreds of years for its outer part. This timescale needs to be compared to the pebble growth timescale (black solid line) and the timescale over which the pebble would be advected by the radial gas flow (red solid line). The shortest of these timescales determines which process dominates. The inner part of the CPD is dominated by the radial drift, while the outer part is dominated by the advection by gas flow. However, there is a narrow region corresponding to the location where the radial gas flow changes direction (see Figure~\ref{fig:infall}), where dust grow should win. This is where we can expect efficient growth and retention of dust (dust trap) which can potentially form satellitesimals via the streaming instability. Satellitesimal formation via the streaming instability should also be very fast, the typical timescales inferred from the hydrodynamical simulations performed in the context of protoplanetary disks are on the order of tens of local orbital timescales \citep[see, e.g.,][]{2016ApJ...822...55S}, which would translate to roughly 1 year at the location of the dust trap. 

\section{Methods}\label{methods}

\subsection{A simplified model}

Numerical models incorporating both detailed growth and advection of dust are computationally expensive as the dust coagulation physics itself is very complex. In this section, we build a simplified model, only including necessary physics in a semi-analytical manner. With such a model, we will gain overview of dust evolution in the complicated environment of circumplanetary disk and, at the same time, we will be able to perform a parameter study as the computational cost of one model is relatively low. 

Dust advection is inseparably connected to the growth physics, as it depends on the dimensionless stopping time $St$ (see Equations \ref{vdrift} and \ref{vadv}). The stopping time is then in turn connected to the physical size of dust grains $a$. Depending on whether this size is smaller or larger than the mean free path of gas molecules $\lambda_{\rm{mfp}}$ (with a factor of 4/9), which corresponds to the change of the drag regimes between the Epstein and Stokes drag, the dimensionless stopping time $St$ can be expressed as
\begin{equation}\label{sizeSt}
{\rm St} = 
\begin{cases}
    \rm{St}_{\rm{Ep}}=\frac{\pi}{2}\frac{a \rho_{\bullet}}{\Sigma_{\rm g}}, \text{if } a < \frac{4}{9}\lambda_{\rm{mfp}},
    \\
    \rm{St}_{\rm{Ep}} \cdot \frac{4}{9} \frac{a}{\lambda_{\rm{mfp}}}, \text{if } a \geq \frac{4}{9}\lambda_{\rm{mfp}},
\end{cases}\end{equation}
where $\rho_{\bullet}$ is the internal density of dust grains.

To avoid direct modeling of dust collisions, we prescribe their growth and fragmentation using a recipe inspired by the work of \citet{2012A&A...539A.148B}. 
Basing on local conditions, we calculate maximum size (or rather the dimensionless stopping time $St$) of dust population at every location in the CPD. We assume that the dust size distribution is fully described by two populations: the small one, corresponding to the initial size of dust grains $a_0$, and the large one, corresponding to the maximum possible size. This maximum size is restricted by the process that dominates dust evolution at a given location. \citet{2012A&A...539A.148B} considered four processes: the initial growth, fragmentation because of turbulence, fragmentation because of differential drift, and the loss of the largest grains due to the radial drift. In this paper, we additionally consider removal of small grains by gas advection. Also, we update the fragmentation because of differential drift regime to take into account the collective drift effect. 

At the start of the simulation, all the dust is assumed to be at $a_0$ size, which we typically set to $10^{-4}$~cm. Following \citet{2012A&A...539A.148B}, the dust growth is prescribed with
\begin{equation}\label{aini}
a_{\rm ini} = a_0\cdot\exp\left(t Z \Omega_{\rm K}\right),
\end{equation}
where $t$ is time since the start of the simulation and $Z$ is the vertically integrated dust-to-gas ratio. A corresponding Stokes number $\rm{St}_{\rm{ini}}$ is calculated using Eq.~\ref{sizeSt}.

Impact speeds of particles increase with their size. This means that small particles grow at every collision but at some point the impact speeds become too high and lead to destructive collisions instead. By comparing the impact speeds driven by turbulence to the fragmentation velocity threshold $v_{\rm f}$, one can derive maximum dimensionless stopping time that the particles can grow to as
\begin{equation}\label{stfrag}
{\rm St}_{\rm{frag}} = {\rm{f_f}} \frac{v_{\rm{f}}^2}{3\alpha c_{\rm{s}}^2},
\end{equation}
where $c_{\rm s}$ is the sound speed of gas, $\alpha$ is the turbulence strength parameter, $\rm{f_f}=0.37$ is a numerical calibration factor, and we typically set $v_{\rm{f}}=10$ m~s$^{-1}$. Following the work of \citet{2014ApJ...785..101F}, we assume that the turbulence in the CPD is low and set $\alpha=10^{-4}$. 

Another major source of collisions is the differential radial drift. As the drift velocity depends on the dimensionless stopping time, particles with different sizes drift with different speeds. There would be no mutual speed in a case of equal-sized particles, but even for a narrow size distribution the drift velocity difference may be significant. Following \citet{2012A&A...539A.148B}, we assume that a large particle would typically collide with particle of half its Stokes number. The work of \citet{2012A&A...539A.148B} did not include the collective drift effect, so the drift velocity was independent on dust density. \citet{2016A&A...594A.105D} showed that including this effect has a major consequences for redistribution of solids. We found that including this effect has importance not only for the redistribution but also for the growth of dust. As the drift speed decreases with increasing dust concentration, the maximum size that the particles may grow to before they fragment becomes
\begin{equation}\label{stdf}
{\rm St}_{\rm{df}} = \frac{v_{\rm{f}}}{2 \eta v_{\rm{K}}} \left(1 + \epsilon\right)^2,
\end{equation}
where $\epsilon$ is the midplane dust-to-gas ratio, $\epsilon=\rho_{\rm d}/\rho_{\rm g}$. The midplane densities of dust and gas are calculated assuming the local hydrostatic equilibrium:
\begin{equation}\label{rhog}
\rho_{\rm g} = \frac{\Sigma_{\rm g}}{\sqrt{2\pi}H_{\rm g}},
\end{equation}
\begin{equation}\label{rhod}
\rho_{\rm d} = \frac{\Sigma_{\rm d}}{\sqrt{2\pi}H_{\rm d}},
\end{equation}
where the gas scale height is calculated as $H_{\rm g} = c_s / \Omega_{\rm K}$, and the dust scale height is calculated as \citep{1995Icar..114..237D}
\begin{equation}\label{Hd}
H_{\rm{d}} = H_{\rm{g}} \sqrt{\frac{\alpha}{\alpha+{\bar{St}}}}.
\end{equation}

Since the simplified method does not include an explicit feedback from the advection to the dust size, we have to make sure that it remains valid even when the advection timescale is shorter than the collisional timescale. By comparing the two timescales, \citet{2012A&A...539A.148B} derived the maximum Stokes number the particles can grow to, before they would be removed by the inward drift as
\begin{equation}\label{stdrift}
{\rm St}_{\rm{drift}} = {\rm{f_d}}\ \frac{Z}{2\eta},
\end{equation}
where the numerical calibration factor $\rm{f_d}=0.55$, and $\eta$ is parametrizing the maximum drift speed (see Eq.~\ref{eta}). 

In the CPD, the radial gas flow resulting from the disk dynamics is significant (see the upper panel of Figure \ref{fig:infall}). Thus, we must also consider the possibility that dust of some size will be carried with the gas flow. Unlike in the case of radial drift, the coupling to the gas is the stronger the smaller the particle size. By comparing the growth timescale and the timescale of dust removal by gas, we derived
\begin{equation}\label{gasadv}
{\rm St}_{\rm{gasadv}}^2 = \frac{v_{\rm{r,gas}}}{Z v_{\rm{K}}} - 1,
\end{equation}
the {\it minimum} Stokes number of grains that are "safe" against the gas flow. This means, that particles with ${St}^2 < {\rm St}_{\rm{gasadv}}^2$ will be removed from their current location by the gas flow. To account for this in the code, we compare the maximum Stokes number at given location $\rm{St}_1$ obtained as the minimum of $\rm{St}_{\rm{ini}}$, ${\rm St}_{\rm{frag}}$, ${\rm St}_{\rm{df}}$, and ${\rm St}_{\rm{drift}}$ to ${\rm St}_{\rm{gasadv}}$ and if $\rm{St}_1^2 < {\rm St}_{\rm{gasadv}}^2$, we set the size of particles at that location back at their starting size $a_0$.

Taking into account the size obtained from the model described above, we track the evolution of dust population and the creation of satellitesimals. The surface density of dust $\Sigma_{\rm d}$ is obtained by solving the advection-diffusion equation
\begin{equation}\label{advdiff}
\frac{\partial \Sigma_{\rm d}}{\partial t} = \frac{1}{r} \frac{\partial}{\partial r}\left[r\left(D_{\rm g}\Sigma_{\rm g}\frac{\partial}{\partial r}\left(\frac{\Sigma_{\rm d}}{\Sigma_{\rm g}}\right)- \Sigma_{\rm d}\bar{v}\right)\right] - \frac{\partial \Sigma_{\rm{s}}}{\partial t},
\end{equation}
where $D_{\rm g}=\alpha c_s H_{\rm g}$ is the diffusion coefficient of gas, $\bar{v}$ is the mass weighted average advection velocity of dust, and $\Sigma_{\rm{s}}$ is the surface density of satellitesimals. The satellitesimals may be formed by the streaming instability, if the conditions proposed by \citet{2014A&A...572A..78D} are fulfilled, namely the midplane density ratio of pebbles with $St > 10^{-2}$ to gas exceeds unity:
\begin{equation}\label{satform}
\frac{\partial \Sigma_{\rm{s}}}{\partial t} = 
\begin{cases}
    \zeta\ \Sigma_{\rm d}\ \Omega_{\rm K}, \text{if } {\rho_{\rm d (St>10^{-2})}/{\rho_{\rm g}}} > 1,
    \\
    0,                                \text{otherwise}
\end{cases}\end{equation}
where $\Omega_{\rm K}$ is the Keplerian frequency and $\zeta$ is the efficiency with which pebbles are turned to satellitesimals. Following \citet{2017A&A...608A..92D}, we use $\zeta=10^{-3}$. 

The advection velocity is calculated as \citep{2016A&A...596L...3I, 2017A&A...602A..21S}:
\begin{equation}\label{vbar}
\bar{v} = \frac{2\eta v_{\rm K} \bar{\rm St} + v_{\rm{r,gas}} \left(1+\epsilon\right)}{{\bar{\rm St}}^2 + \left(1+\epsilon\right)^2},
\end{equation}
which connects the two contributions: the radial drift caused by loss of angular momentum because of the headwind (see Eq.~\ref{vdrift}) and the advection by gas flow (Eq.~\ref{vadv}). Additionally, we include the effect of collective drift, which means that the advection velocity decreases as the dust concentration increases. The collective drift is taken into account with the midplane dust-to-gas ratio $\epsilon=\rho_{\rm d}/\rho_{\rm g}$.

The advection of dust is performed by taking into account the mass-weighted average velocity $\bar{v}$ which relies on the mass-weighted average Stokes number of the dust population $\bar{\rm St}$. This is calculated based of the estimated dust distribution (see \citealt{2012A&A...539A.148B}). If the growth is limited by fragmentation, we assume that 75\% of dust surface density corresponds to the largest grains and the rest to the smallest grains, while if the growth is limited by drift, 97\% of dust surface density corresponds to the largest grains.

In the simplified model, we keep the CPD gas and temperature profile fixed. Since all the dust evolution timescales involved are much shorter than the CPD dispersal timescale, which should similar to the dispersal of protoplanetary disk (on the order of 1 to 10~Myrs, see e.g.~\citealt{2001ApJ...553L.153H}), it seems to be a justified approach.

\subsection{Monte Carlo simulation}

\begin{figure*}
\plotone{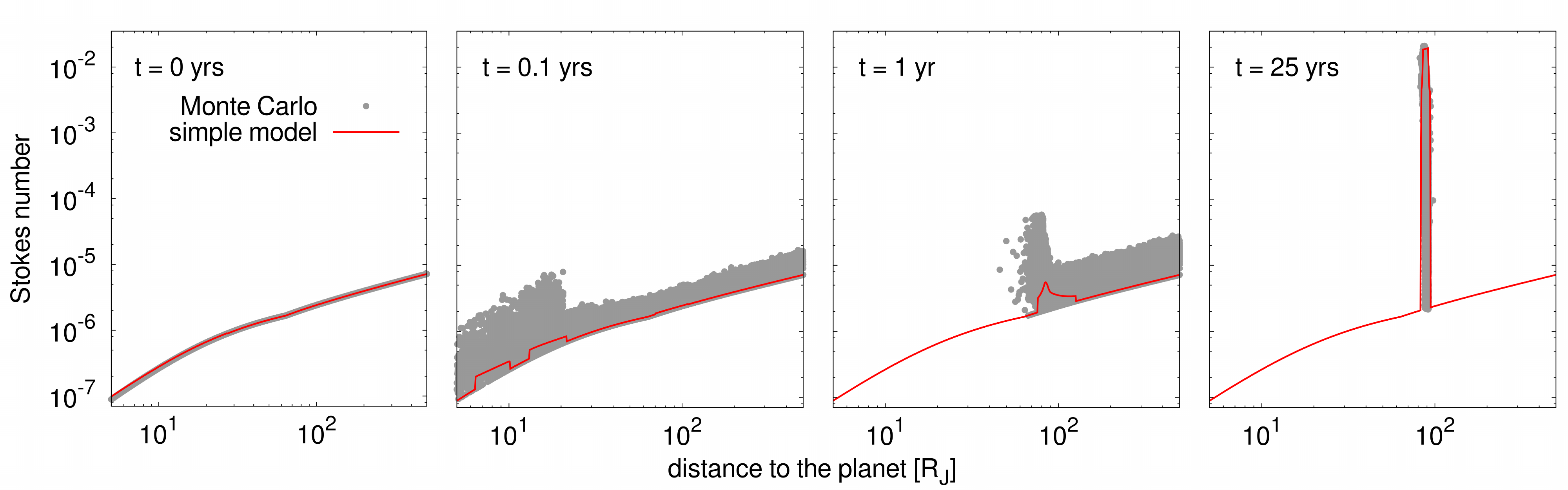}
\caption{Stokes number of dust particles obtained in the Monte Carlo simulations (points) and in the simplified model (red solid line) at different stages of evolution.}\label{fig:MCsizes}
\end{figure*}

To confirm the validity of our simple model described in the previous section, we ran a 2-D ($r+z$) Monte Carlo simulation using the code introduced by~\citet{2013A&A...556A..37D}. This Lagrangian code is based on the representative particle approach \citep{2008A&A...489..931Z}, in which the total mass of dust is divided into a limited number of representative bodies, each of them defined by a set of identical physical particles. We follow the interactions of the representative particles with the physical particles reproduced by their counterparts. This is a valid approach since the interactions between two representative particles are unlikely as the number of such particles is much lower than the number of physical bodies. To account for the local nature of collisions, we divide the disk into grid cells using an adaptive grid algorithm and allow for collisions only between particles placed in the same grid cell. We refer the interested reader to \citet{2013A&A...556A..37D} for more detailed description of the method.

We adopted exactly the same CPD model as described above and assumed the starting dust-to-gas ratio of 0.01. For the Monte Carlo test, the dust infall on the disk was excluded. We started the Monte Carlo simulation with 512,000 representative particles distributed such that the surface density of dust has the same profile as the surface density of gas. The particles are initially distributed between 5~R$_{\rm J}$ and 700~$R_{\rm J}$ from the central planet. We assumed that the initial size of dust grains is $a_0=10^{-4}$~cm and that they have the internal density of $\rho_{\bullet}=3$~g~cm$^{-3}$.

The Monte Carlo code follows the advection of dust particles due to vertical settling and radial drift, their collisions, and accounts for the possibility of satellitesimals formation via streaming instability. Five sources of collision velocities were taken into account: the Brownian motions, turbulence (with formulas derived by \citealt{2007A&A...466..413O}), vertical settling, and the differential azimuthal and radial drift. The collective drift effects were taken into account when calculating the contributions from azimuthal and radial drift with the formulas derived by \citet[][their Eqs. 48-49]{2012ApJ...752..106O}. Collisions with the impact velocity $\Delta{v}$ lower than the threshold velocity $v_{\rm f}=10$~m~s$^{-1}$ result in sticking and collisions with $\Delta{v} > v_{\rm f}$ result in fragmentation of the dust aggregates. 

Satellitesimal formation via streaming instability was included with the same algorithm as used in \citet{2014A&A...572A..78D}. If the conditions for streaming instability are fulfilled (see Eq.~\ref{satform}), representative particles corresponding to the largest dust aggregates are removed from the simulation and their mass is added to the satellitesimal reservoir. There is no further satellitesimal evolution included in the current version of the code.

The Monte Carlo code has an advantage of being very flexible as it does not make many ad-hoc assumptions, like it is in the case of our simplified model. Thanks to this, it is a perfect tool to explore the dust evolution in a relatively exotic environment of the CPD and enables us to validate the simplified model. However, the numerical cost of the Monte Carlo simulation is very high. It took about 3,800 CPU hours to evolve the system through 7,000 years, while for the simple model it would be a cost of only about 7 CPU hours. 

\section{Results}\label{results}

\subsection{Comparison between the Monte Carlo simulations and the simplified model}

In this Section, we present the results of our numerical models.
First, we focus on validating the simplified method by comparing its results to the outcome of Monte Carlo run. Due to numerical limitations, this models do not include the continuous infall of dust from the protoplanetary disk to the CPD. 

Figure~\ref{fig:MCsizes} shows the comparison between dimensionless stopping time of the representative particles in the Monte Carlo simulation and predicted by the simplified model, at the different stages of evolution. At the beginning of the simulation, all grains have a size of $a_0=10^{-4}$~cm, corresponding to the dimensionless stopping time of $St=10^{-7}$ at the inner edge of the CPD and $St=10^{-5}$ at its outer edge. The growth proceeds inside-out, as the growth timescale is the shortest at the inner edge of the disk (see Figure~\ref{fig:times}). However, as the radial drift and removal of dust by gas advection have similar timescales, already after 1 year of evolution the inner part of the disk is significantly depleted. The remaining dust gathers in the region where the radial gas flow changes direction from outward to inward, and this is the only location where dust can grow to pebble sizes ($St\geq10^{-2}$). The rightmost panel of Figure~\ref{fig:MCsizes} corresponds to a steady state, where all the dust particles are gathered in the trap caused by the gas flow structure. Due to the short timescales, the steady state is obtained already after $\sim25$~years of evolution.

\begin{figure}
\plotone{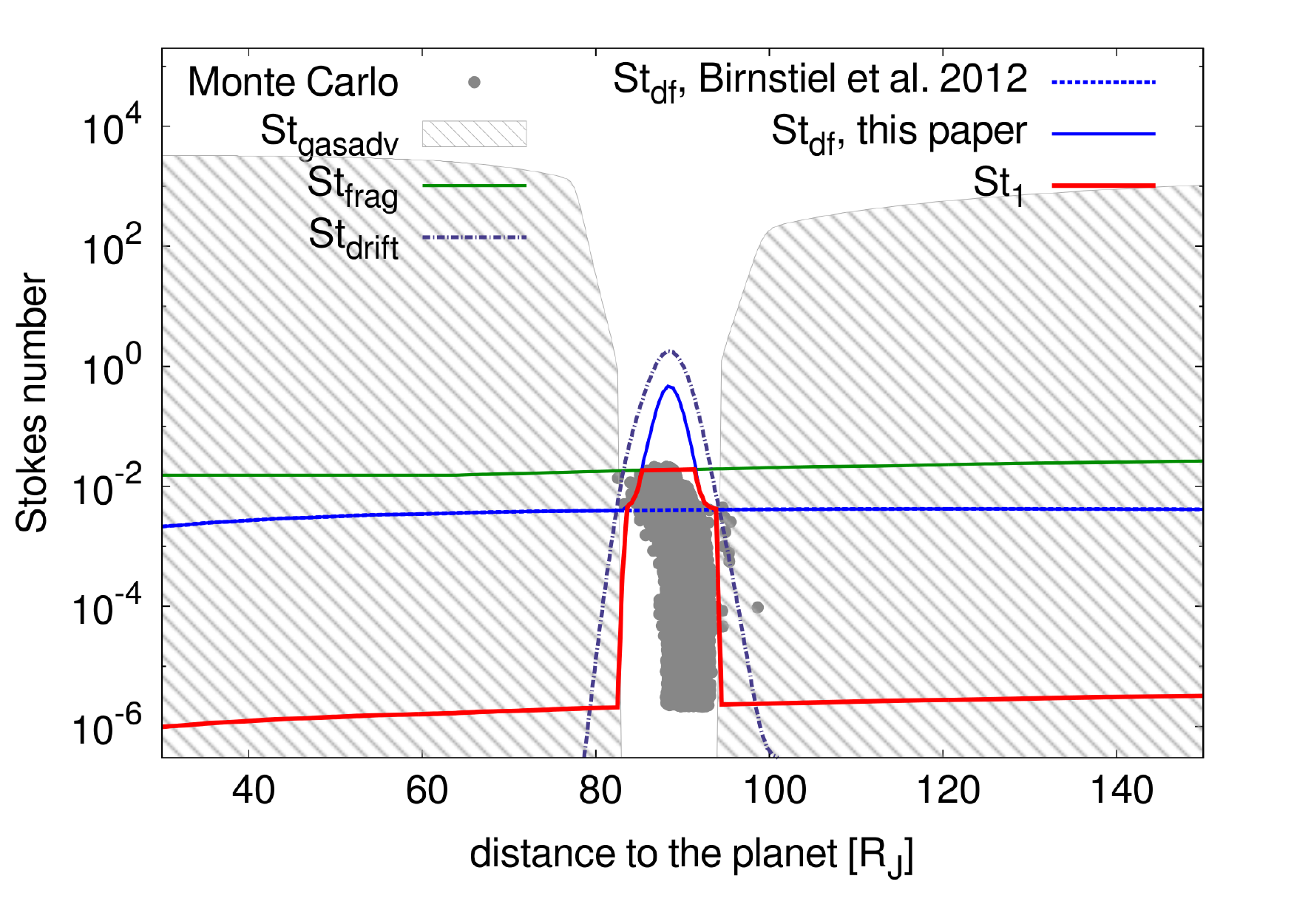}
\caption{Comparison of the dust Stokes number in the vicinity of the dust trap obtained in the Monte Carlo simulation (points) and in the simplified model (red solid line). Additionally, the maximum Stokes number profiles calculated in the simplified model while taking into account different processes are plotted. The hatched region shows where the removal of dust by the gas flow is efficient.}\label{fig:sizesexact}
\end{figure}

A zoom onto to the dust trap region is provided in Figure~\ref{fig:sizesexact}. The sizes outside of the trap are controlled by removal of dust by the gas advection (the hatched region, ${\rm St}_{\rm gasadv}$, see Eq.~\ref{gasadv}). Inside of the trap, the maximum Stokes number is defined by fragmentation driven by turbulence (green solid line, ${\rm St}_{\rm frag}$, Eq.~\ref{stfrag}) and fragmentation driven by the differential drift (blue solid line, ${\rm St}_{\rm df}$, Eq.~\ref{stdf}). Importance of taking into account the collective drift effect when calculating the maximum size due to the differential drift is highlighted here. If the collective drift is not taken into account (blue dotted line), the size coming from the simple model is significantly smaller than obtained with the Monte Carlo code and what is more, the grains would be too small to trigger the streaming instability. 

In general, we obtain a reasonably good agreement between the two methods concerning the global evolution pattern and the dust sizes. In the following sections, we focus on the simulations done with the simplified model, as due to the high computational cost of Monte Carlo method, we could only perform a few tests with it.

\subsection{Models with infall}

Due to limitations of the Monte Carlo method, the infall of dust onto the CPD was excluded in the previously described runs. We included it in the framework of the simplified method. We assumed that the dust infall is described with the same profile as the gas infall measured in the hydrodynamical simulations (see the bottom panel of Figure~\ref{fig:infall}), such that 
\begin{equation}
\dot{\Sigma}_{\rm d,infall} = Z_0\cdot\dot{\Sigma}_{\rm g,infall},
\end{equation}
where $Z_0$ is the initial, global dust-to-gas ratio. We assume that the infalling dust has size of $a_0$. 

\begin{figure*}
\plotone{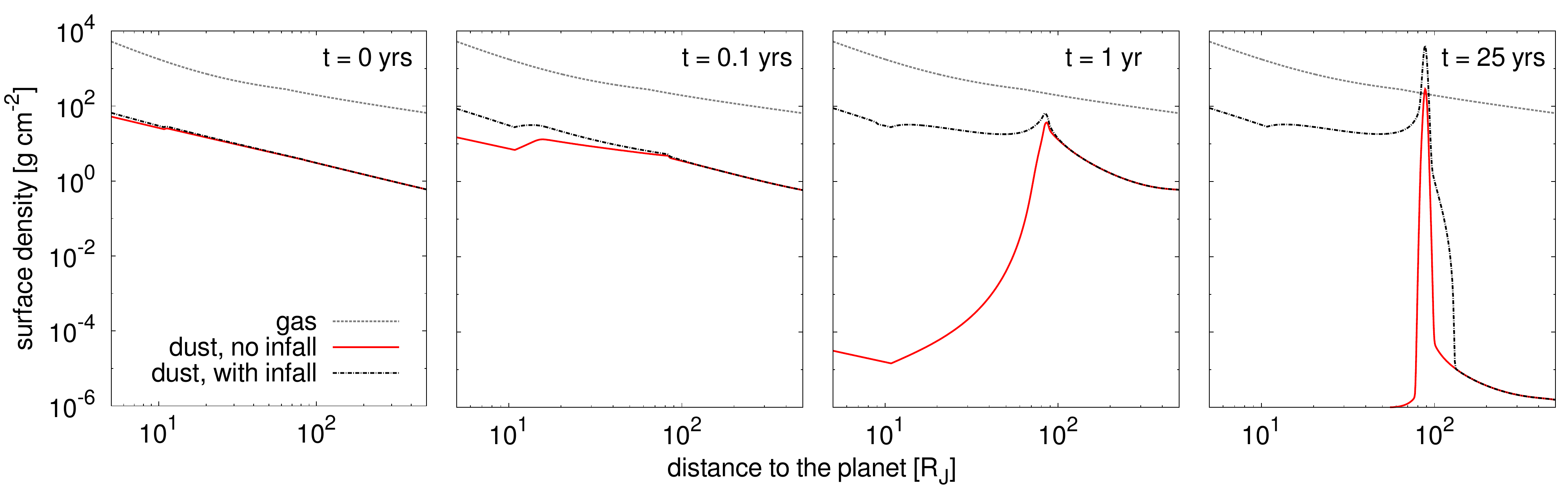}
\caption{Evolution of the surface density of dust in the simple models with and without dust infall and with $Z_0=0.01$. The (constant) density of gas is displayed for reference.}\label{fig:surfdeevo}
\end{figure*}
Figure~\ref{fig:surfdeevo} presents the comparison of dust surface density evolution obtained in the models without infall and with infall for the default value of $Z_0=10^{-2}$. In the model without infall, the inner part of CPD is dust depleted after just 1 year, corresponding to the dust evolution timescale close to the planet (see Figure~\ref{fig:times}). In the model with infall, this depletion is hindered and the surface density inside of the dust trap reaches higher values, as the dust population is constantly replenished.

In the case without infall, the dust reservoir is limited and all the dust is either lost due to the radial drift or gathers in the trap region. This dust then grows to pebble sizes and its size distribution is regulated by the coagulation-fragmentation equilibrium. The pebbles are transferred to satellitesimals via the steaming instability, but at some point there is not enough dust to support further satellitesimal formation. In the infall case, the dust reservoir is constantly refilled and a steady-state profile is formed when the infall, advection, diffusion, and satellitesimal formation balance, and the dust mass in the CPD stays constant. The CPD acts as a satellitesimal factory, quickly processing the fresh dust, constantly delivered from the circumstellar disk, to pebbles and satellitesimals.

\begin{figure}
\plotone{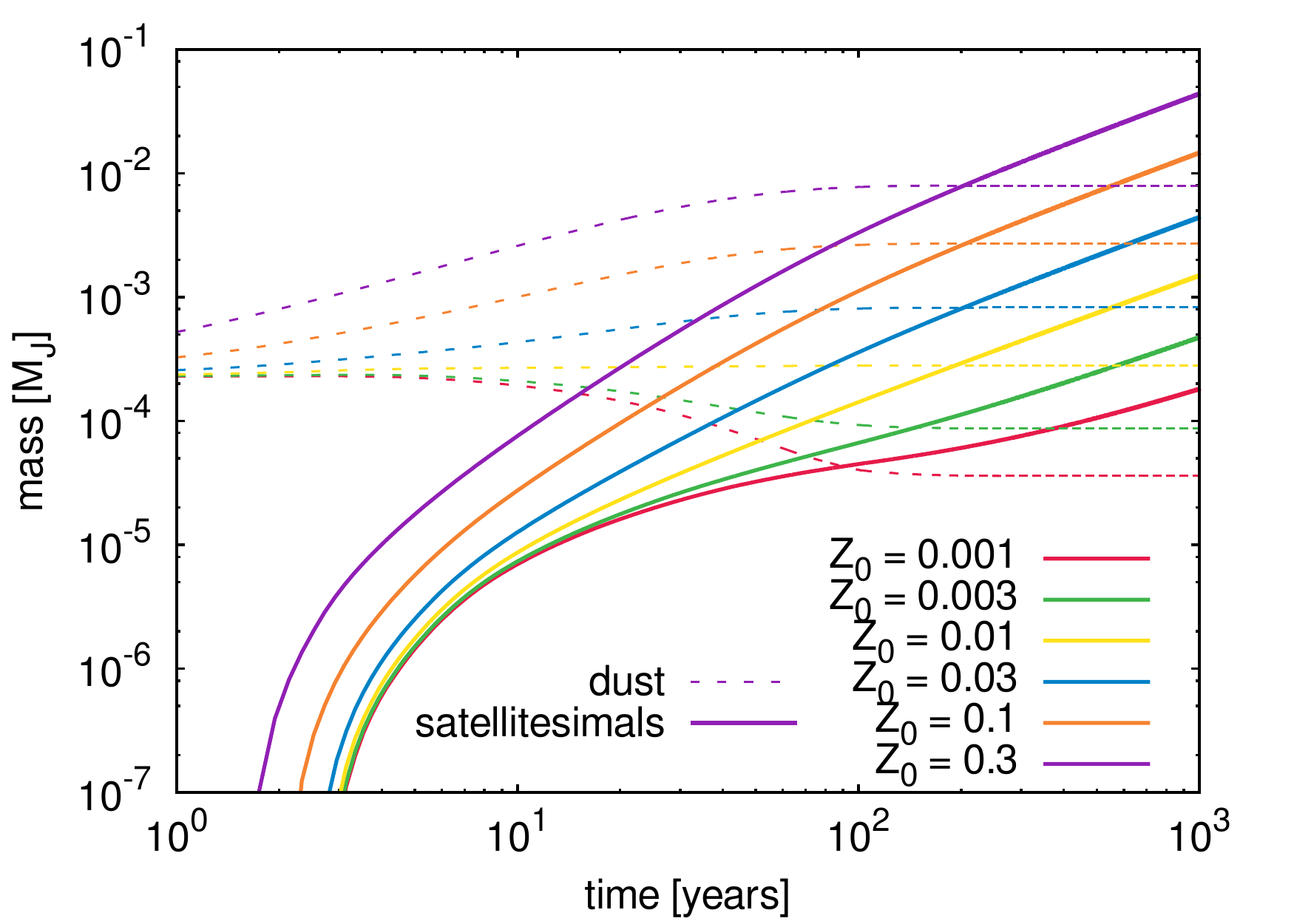}
\caption{Time evolution of dust and satellitesimals mass reservoirs in the simplified models with different initial dust-to-gas ratio $Z_0$.}\label{fig:massevo} 
\end{figure}

Figure~\ref{fig:massevo} presents the time evolution of the CPD mass reservoir for models with different dust-to-gas ratio $Z_0$. All the models arrive at their steady-state well before 1,000~years of evolution, however it takes a little bit longer for models with lower $Z_0$. The final mass of dust and the satellitesimal formation rate are scaling linearly with $Z_0$. We measure the satellitesimal formation rate as 
\begin{equation}
\dot{M}_{\rm satellitesimals} = \left(\frac{Z_0}{0.01}\right)\cdot2.4\cdot10^{-7}\ \frac{{\rm M}_{\rm J}}{\rm year}.
\end{equation}
This formation rate of satellitesimals may seem high as it would only take about 1,000~years to form enough satellitesimals to reproduce all the Galilean satellites (for $Z_0=0.01$). However, as showed by \citet{2018MNRAS.480.4355C}, majority of the satellitesimals will be lost into the central planet due to their fast radial migration. 

\begin{figure}
\plotone{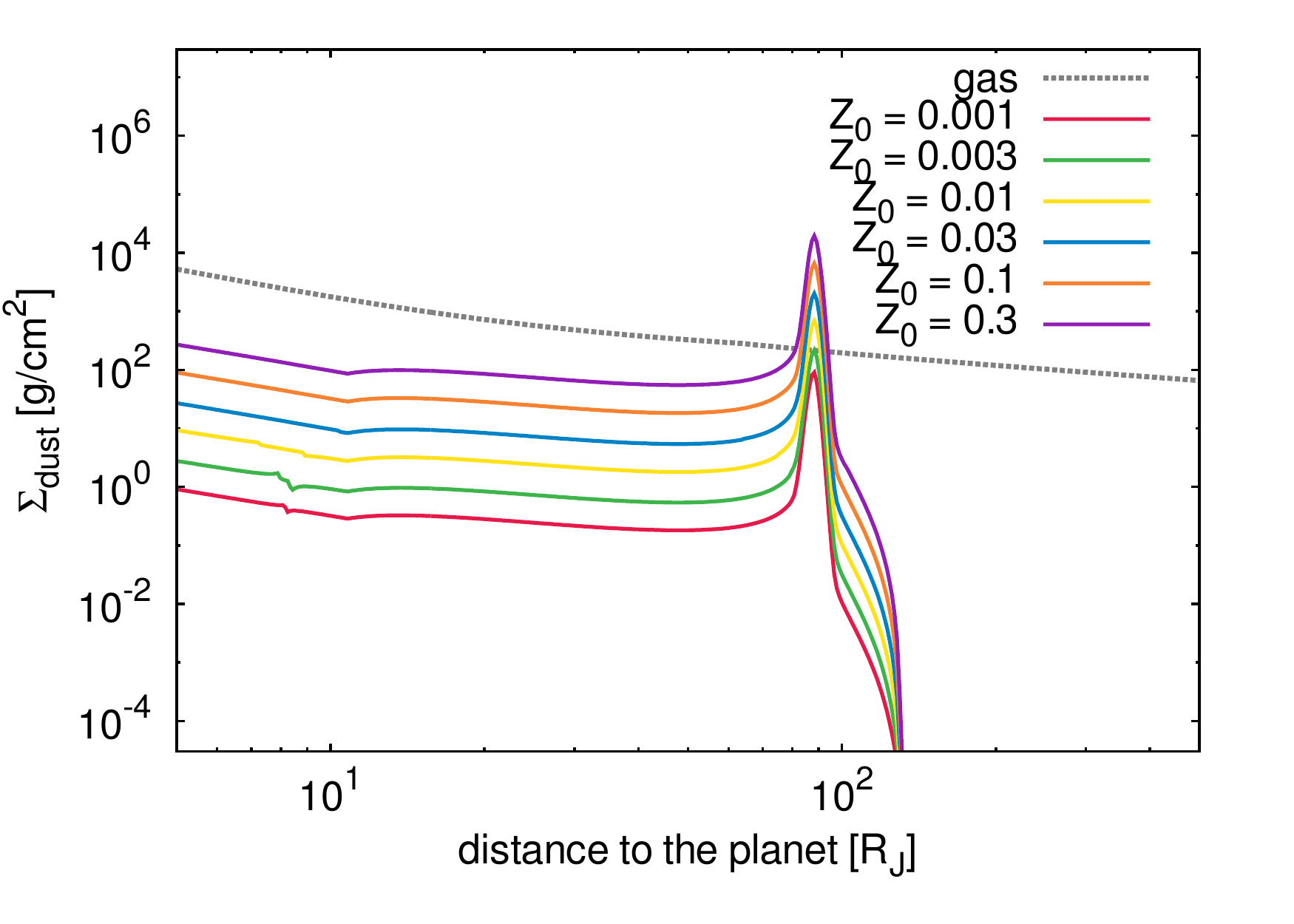}
\caption{Steady-state surface density of dust profiles obtained in the simulations with different initial solids-to-gas ratio $Z_0$.}  \label{fig:surfdefinal}
\end{figure}

Similarly to the satellitesimal formation rate, the steady-state surface density of dust displayed in Figure~\ref{fig:surfdefinal} scales linearly with $Z_0$. This profile results from an equilibrium between the dust infall, advection and diffusion, and satellitesimal formation via the streaming instability and is peaked around the dust trap region, at about 85~R$_{\rm J}$ from the planet. These steady-state profiles are obtained very quickly, in all the cases before 100 years of evolution. 

\begin{figure}
\plotone{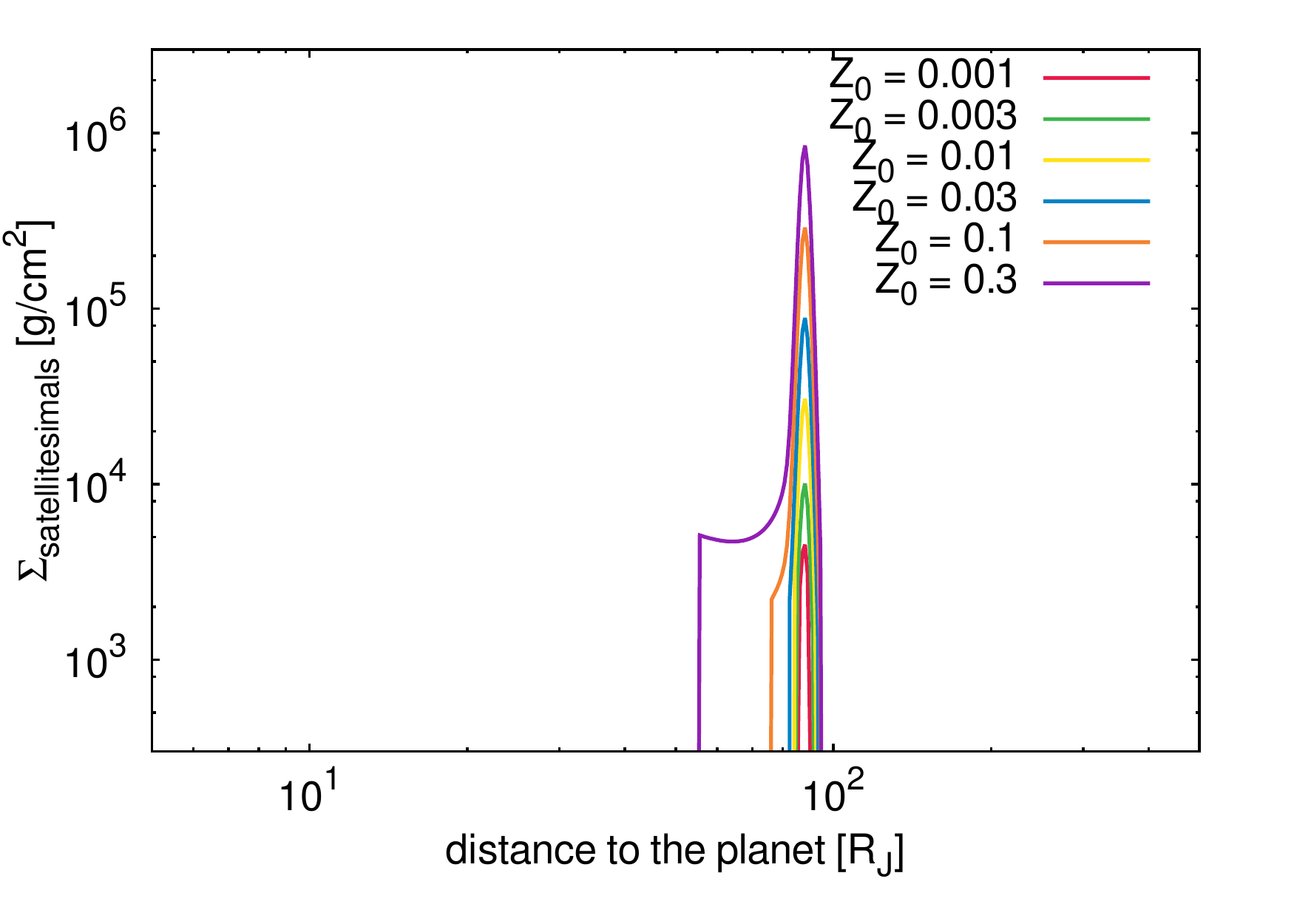}
\caption{Surface density of satellitesimals obtained in the simulations with different initial solids-to-gas ratio after 1,000 years of evolution.}\label{fig:surfdesat}
\end{figure}

Figure~\ref{fig:surfdesat} presents the surface density of satellitesimals obtained in the simplified models with different starting dust-to-gas ratio $Z_0$. In all cases, most of the satellitesimals form directly in the dust trap region. In the models with the highest $Z_0$, the satellitesimal formation region is a little bit wider and extends inwards from the trap. This is because with more dust, the streaming instability is already active before all the dust is gathered in the trap region (see the evolution showed in Figures~\ref{fig:MCsizes} and \ref{fig:surfdeevo}), before the the steady-state profile is reached.

\section{Discussion}\label{discussion}

We found that the key feature that stops dust particles form falling onto the central planet and enables satellitesimal formation within the CPD is existence of the outward gas flow region. The fact that the midplane is dominated by outward gas flow was also pointed out by other hydrodynamical simulations performed with different numerical codes \citep{2010MNRAS.405.1227M, 2012ApJ...747...47T, 2016ApJ...832..105F}, so this feature of the CPD seems robust. However, the location at which this outward flow changes back to inwards flow, which determines where the dust trap and satellitesimals form, depends sensitively on parameters used in the numerical models, particularly on viscosity as shown by \citet{2014ApJ...782...65S}, but also on the disk opacity, infall of matter from the circumstellar disk, and mass and temperature of the central planet. We focused our work on the CPD model extracted from one of the simulations presented by \citet{2017ApJ...842..103S}, where this trap is at 85~R$_{\rm J}$ from the planet.

The uncertainty on the trap location may pose a question of whether the dust trap and streaming instability scenario will still work in other CPD models. One may imagine that if, for example, the dust infall region is much closer to the planet (see Figure~\ref{fig:infall}), most of the dust would be lost because the inward drift and the dust trap would be inefficient. However, since the infall of material is naturally connected to the outward flow region (i.e.~the presence of meridional circulation between the CPD and the circumstellar disk, see e.g.~\citealt{2012ApJ...747...47T, 2014ApJ...782...65S,2016ApJ...832..105F}), we have a good reason expect that the scenario presented in this paper is in fact universal and will work in every circumplanetary disk.

Typical timescale of the models presented in this paper is only 1,000 years. We found that this is still much longer than the dust needs to achieve its equilibrium profile. In fact, due to extremely short evolution timescales (see Figure~\ref{fig:times}), dust pile-up region forms very quickly. In our models, the gas disk structure is fixed. However, in reality, as it is sensitive to many parameters such as the temperature profile and infall form the circumstellar disk, we expect that the gas flow structure may evolve on timescales comparable to the circumstellar disk evolution. This would mean that dust would be able to adjust to changing location of the trap and keep forming satellitesimals as long as there is enough new material falling from the circumstellar disk and the outward gas flow region still exists. This is indeed what we expect, basing on the work of \citet{2017ApJ...842..103S}, where snapshots of the evolving planet-disk system were produced by assuming that the planet is cooling over time. They found that the general meridional flow structure is kept as the planet cools down.

To address the problem of long term evolution of the CPD, in \citet{2018MNRAS.480.4355C} we assumed an exponential dispersal of the CPD and postulated that the dust surface density is reduced accordingly while keeping its profile. To test this assumption, we ran a model with the gas surface density and infall profile reduced by 50\%. The dust profile that we obtained is the same shape as the equilibrium profile displayed in Figure~\ref{fig:surfdefinal}, but also reduced by 50\%. The satellitesimal formation rate decreases accordingly. Thus, we expect that the the slow dispersal of CPD does not change the dust profile shape, but only reduces the amount of dust present and the satellitesimal formation rate. 

Performing a very long timescale runs that would explicitly include the CPD dispersal and cooling is beyond the scope of this paper. However, we must acknowledge that the CPD simulation used in this paper is too hot to support the existence of water ice at the present time, while the Galilean satellites in fact contain significant amount of the water ice \citep{1999Sci...296...77S}. This means that they had to form late in the CPD evolution, when it was cold enough for the water ice to exist, as it was already pointed out by, e.g., \citet{2015ApJ...806..181H}, and they could not have been significantly heated after their formation \citep{2015A&A...579L...4H}. The entire evolution of CPD cannot be covered by the expensive hydrodynamical simulations. However, we did follow the long-term evolution in a 1-D semi-analytical model (see \citealt{2018MNRAS.480.4355C}), which showed that in the last few hundred thousand years before the disk dissipates, the CPD is cool enough to produce icy satellitesimals, and this timescale is long enough to form at least a few generation of satellites. On a side note, the snow lines could cause additional modifications to the CPD structure. Similarly as in the case of a circumstellar disk, sharp opacity transitions related to snow lines could lead to development of dust traps \citep[see, e.g.][]{2007ApJ...664L..55K, 2008A&A...487L...1B}, which could open a possibility to satellitesimal formation at more than one location.

In our models, we used a set of default parameters, such as the infalling dust size of $a_0=10^{-4}$~cm and the threshold fragmentation velocity $v_{\rm f}=10$~cm~s$^{-1}$. We tested that varying the $a_0$ between $10^{-5}$~cm and $10^{-3}$~cm does not impact our results. This is because dust of all sizes is quickly transported to the dust trap region, where it can grow to the cm-sizes corresponding to $St\approx10^{-2}$ (see Figure~\ref{fig:sizesexact}). Infall of even larger grains seems unrealistic, as the Jupiter mass planet opens a gap in the protoplanetary disk that acts as a pressure bump and stops larger dust aggregates \citep{ 2012A&A...545A..81P}. 

Some of the previous research suggested that planetesimals will pass through the pressure bump of the planetary gap and get captured by the CPD, providing a reservoir for satellite seeds \citep{2007ApJ...666..447Z,2008ApJ...684.1416S,2014ApJ...784..109T,2015ApJ...806..203D,2016ApJ...820..128S,2017ApJ...839...66S,2018AJ....155..224R}. Even if this mechanism might be possible, as we showed in this work, the satellitesimals can efficiently form within the CPD and there is no need for the external source of the satellite formation seeds.

Our results are a bit more sensitive to fragmentation threshold velocity $v_{\rm f}$. The maximum size of dust that can grow in the trap region is determined by turbulence induced fragmentation, which is very sensitive to the value of $v_{\rm f}$ (see Eq.~\ref{stfrag}). We found that with a low fragmentation velocity $v_{\rm f}<8$~m~s$^{-1}$, no pebbles with $St>10^{-2}$ would grow in the trap region and, consequently, no satellitesimals would be formed by the streaming instability. Laboratory experiments performed for silicate grains indicated fragmentation threshold velocities on the order of 1~m~s$^{-1}$ \citep{2010A&A...513A..56G}. This $v_{\rm f}$ is expected to be higher for porous grains \citep{2011ApJ...737...36W} and possibly also for organic materials \citep{2016Icar..267..154P}. It is well established that water ice grains are significantly more sticky than silicates, with $v_{\rm f}$ between 10~m~s$^{-1}$ and 30~m~s$^{-1}$ \citep{2009ApJ...702.1490W, 2014MNRAS.437..690A, 2015ApJ...798...34G}. In the models presented in this paper, we applied $v_{\rm f}=10$~m~s$^{-1}$. For even higher values of $v_{\rm f} \gtrsim 70$~m~s$^{-1}$, fragmentation would not happen at all, enabling direct growth to satellitesimal sizes in the dust trap region. Outside of the trap region, particles would still be removed by the radial drift or gas advection faster than they could grow.

In our models, we assumed that after reaching a critical dust-to-gas ratio in the dust trap, the pebbles are transformed into satellitesimals via the streaming instability. However, the feasibility of the streaming instability in the CPD environment is yet to be tested. The existing models were only performed in the context of the circumstellar disk, and they found that the streaming instability is typically enhancing the dust density $\rho_{\rm d}$ to about $10^3 \cdot \rho_{\rm g}$, which is above the Roche density in the circumstellar disk environment, so that the dust clumps undergo gravitational collapse \citep{2010ApJ...722.1437B}. In our model of the CPD, at the location of the dust trap, the Roche density is on the order of $10^4 \cdot \rho_{\rm g}$, which may not be easily obtained by the streaming instability \citep[see, however, the results of][where dust density of $10^4\cdot \rho_{\rm g}$ is obtained in most of the models]{2016ApJ...822...55S}. However, even if the streaming instability is not operating, taking into account the constant delivery of dust from the circumstellar disk, the satellitesimals would form anyway, either by direct gravitational instability or direct growth.

\section{Summary}\label{summary}

In this paper, we addressed the problem of satellite formation in a circumplanetary disk. We coupled the outcome of state-of-the art hydrodynamic simulations to a dust evolution model and found that the satellite seeds (satellitesimals) may be formed very efficiently but only at one location the circumplanetary disk. The gas flows outwards in a significant part of the disk stopping the radial drift of dust particles and creating a pile-up region -- the dust trap -- where the dust-to-gas ratio is significantly enhanced. The high concentration of solids leads to an efficient growth of dust to pebbles and subsequent formation of gravitationally bound objects via the streaming instability. We showed that the dust evolution and satellitesimal formation is extremely fast, much faster than the projected disk dispersal timescale. The constant feed of dusty material from the circumstellar disk to the circumplanetary disk, and the very short timescale of dust evolution, turns the circumplanetary disk into a satellitesimal factory, continuously processing the infalling dust to pebbles.

Our findings are very important to satellite formation models. In a corresponding paper \citep{2018MNRAS.480.4355C}, we showed that, with the results obtained with our simplified dust model, also the further growth of satellitesimals to satellites is fast, typically about a few tens of thousands years. However, due to their radial migration, majority of these forming satellites are lost to the central planet, enhancing its envelope with heavy elements. Still, in many cases, 3-4 satellites of the last generation that formed will survive when the gas dissipates from the disk (and therefore when the migration stops). In conclusion, the satellitesimal formation scenario presented in this paper enabled \citet{2018MNRAS.480.4355C} to successfully reproduce the Galilean satellites formation within the population synthesis models.

\acknowledgments
The authors thank Shigeru Ida, Satoshi Okuzumi, Chris Ormel, Marco Cilibrasi, and Lucio Mayer for useful discussions. We also thank the referee, Yuhito Shibaike, for comments that helped us to improve the paper. J.D.~acknowledges the support of the National Centre for Competence in Research PlanetS supported by the Swiss National Science Foundation and funding from the European Research Council (ERC) under the European Union’s Horizon 2020 research and innovation programme under grant agreement No 714769. J.Sz.~acknowledges the support from the Swiss National Science Foundation (SNSF) Ambizione grant PZ00P2\_174115. The hydrodynamic computations have been done on the``M\"onch" machine hosted at the Swiss National Computational Centre.

\software{JUPITER code \citep{2016MNRAS.460.2853S}, 2D~Monte Carlo dust evolution code \citep{2013A&A...556A..37D}}

\bibliographystyle{aasjournal}
\bibliography{cpd.bib}

%% This command is needed to show the entire author+affilation list when
%% the collaboration and author truncation commands are used.  It has to
%% go at the end of the manuscript.
%\allauthors

%% Include this line if you are using the \added, \replaced, \deleted
%% commands to see a summary list of all changes at the end of the article.
%\listofchanges

\end{document}